\documentclass[12pt]{article}
\usepackage{amsfonts}
\usepackage{amssymb}%,letterspace}
\usepackage{graphics,psboxit,amsmath}
\usepackage{subfigure}
\usepackage{graphicx}
\usepackage{verbatim}
\usepackage{epic}

\def\hybrid{\topmargin 0pt \oddsidemargin 0pt %%%%%%%%%%%%%% Archive-30pt
        \headheight 0pt \headsep 0pt
        \textwidth 16,0cm % A4 paper
        \textheight 22,5cm % A4 paper
        \marginparwidth .875in
        \parskip 5pt plus 1pt \jot = 1.5ex}

\hybrid

\def\baselinestretch{1.2}

\catcode`\@=11

\def\marginnote#1{}
\newcount\hour
\newcount\minute
\newtoks\amorpm
\hour=\time\divide\hour by60
\minute=\time{\multiply\hour by60 \global\advance\minute by-\hour}
\edef\standardtime{{\ifnum\hour<12 \global\amorpm={am}%
        \else\global\amorpm={pm}\advance\hour by-12 \fi
        \ifnum\hour=0 \hour=12 \fi
        \number\hour:\ifnum\minute<10 0\fi\number\minute\the\amorpm}}
\edef\militarytime{\number\hour:\ifnum\minute<10 0\fi\number\minute}
\def\draftlabel#1{{\@bsphack\if@filesw {\let\thepage\relax
   \xdef\@gtempa{\write\@auxout{\string
      \newlabel{#1}{{\@currentlabel}{\thepage}}}}}\@gtempa
   \if@nobreak \ifvmode\nobreak\fi\fi\fi\@esphack}
        \gdef\@eqnlabel{#1}}
\def\@eqnlabel{}
\def\@vacuum{}
\def\draftmarginnote#1{\marginpar{\raggedright\scriptsize\tt#1}}

\def\draft{\oddsidemargin -.5truein
        \def\@oddfoot{\sl preliminary draft \hfil
        \rm\thepage\hfil\sl\today\quad\militarytime}
        \let\@evenfoot\@oddfoot \overfullrule 3pt
        \let\label=\draftlabel
        \let\marginnote=\draftmarginnote
   \def\@eqnnum{(\theequation)\rlap{\kern\marginparsep\tt\@eqnlabel}%
\global\let\@eqnlabel\@vacuum} }

\def\draft2{
        \def\@oddfoot{\sl preliminary draft \hfil
        \rm\thepage\hfil\sl\today\quad\militarytime}
        \let\@evenfoot\@oddfoot \overfullrule 3pt
        \let\label=\draftlabel
        \let\marginnote=\draftmarginnote
   \def\@eqnnum{(\theequation)\rlap{\kern\marginparsep\tt\@eqnlabel}%
\global\let\@eqnlabel\@vacuum} }

\def\preprint{\twocolumn\sloppy\flushbottom\parindent 2em
        \leftmargini 2em\leftmarginv .5em\leftmarginvi .5em
        \oddsidemargin -.5in \evensidemargin -.5in
        \columnsep .4in \footheight 0pt
        \textwidth 10.in \topmargin -.4in
        \headheight 12pt \topskip .4in
        \textheight 6.9in \footskip 0pt
        \def\@oddhead{\thepage\hfil\addtocounter{page}{1}\thepage}
        \let\@evenhead\@oddhead \def\@oddfoot{} \def\@evenfoot{} }

\def\numberbysection{\@addtoreset{equation}{section}
        \def\theequation{\thesection.\arabic{equation}}}

\def\underline#1{\relax\ifmmode\@@underline#1\else
        $\@@underline{\hbox{#1}}$\relax\fi}

\def\titlepage{\@restonecolfalse\if@twocolumn\@restonecoltrue\onecolumn
     \else \newpage \fi \thispagestyle{empty}\c@page\z@
        \def\thefootnote{\fnsymbol{footnote}} }

\def\endtitlepage{\if@restonecol\twocolumn \else \newpage \fi
        \def\thefootnote{\arabic{footnote}}
        \setcounter{footnote}{0}} %\c@footnote\z@ }

\catcode`@=12
\relax

\def\figcap{\section*{Figure Captions\markboth
        {FIGURECAPTIONS}{FIGURECAPTIONS}}\list
        {Figure \arabic{enumi}:\hfill}{\settowidth\labelwidth{Figure
999:}
        \leftmargin\labelwidth
        \advance\leftmargin\labelsep\usecounter{enumi}}}
 \relax
\def\tablecap{\section*{Table Captions\markboth
        {TABLECAPTIONS}{TABLECAPTIONS}}\list
        {Table \arabic{enumi}:\hfill}{\settowidth\labelwidth{Table
999:}
        \leftmargin\labelwidth
        \advance\leftmargin\labelsep\usecounter{enumi}}}
 \relax
\def\reflist{\section*{References\markboth
        {REFLIST}{REFLIST}}\list
        {[\arabic{enumi}]\hfill}{\settowidth\labelwidth{[999]}
        \leftmargin\labelwidth
        \advance\leftmargin\labelsep\usecounter{enumi}}}
 \relax
\makeatletter
\newcounter{pubctr}
\def\publist{\@ifnextchar[{\@publist}{\@@publist}}
\def\@publist[#1]{\list
        {[\arabic{pubctr}]\hfill}{\settowidth\labelwidth{[999]}
        \leftmargin\labelwidth
        \advance\leftmargin\labelsep
        \@nmbrlisttrue\def\@listctr{pubctr}
        \setcounter{pubctr}{#1}\addtocounter{pubctr}{-1}}}
\def\@@publist{\list
        {[\arabic{pubctr}]\hfill}{\settowidth\labelwidth{[999]}
        \leftmargin\labelwidth
        \advance\leftmargin\labelsep
        \@nmbrlisttrue\def\@listctr{pubctr}}}
 \relax
\makeatother

\def\ba{\begin{equation}}
\def\ea{\end{equation}}
\def\no{\noindent}

\def\IR{\relax{\rm I\kern-.18em R}}

\begin{document}
%\draft2

%\renewcommand{\theequation}{\arabic{equation}}
%\renewcommand{\theequation}{\thesection.\arabic{equation}}

\renewcommand{\theequation}{\thesection.\arabic{equation}}
\csname @addtoreset\endcsname{equation}{section}

\newcommand{\eqn}[1]{(\ref{#1})}
\newcommand{\be}{\begin{eqnarray}}
\newcommand{\ee}{\end{eqnarray}}
\newcommand{\non}{\nonumber}
\begin{titlepage}
\strut\hfill
\vskip 1.3cm
\begin{center}

%\begin{center}

{\bf {\Large New reflection matrices for the $U_q(gl({\mathrm m}|{\mathrm n}))$ case %as super representations of the affine Hecke
algebra }}

\vspace{0.5in}

{\bf Anastasia Doikou}  and {\bf Nikos Karaiskos}

%\vspace{10mm}
 \vskip 0.02in

 {\footnotesize  University of Patras, Department of Engineering Sciences,\\ GR-26500 Patras, Greece}\\[2mm]
\noindent
{\footnotesize {\tt adoikou$@$upatras.gr, nkaraiskos@upatras.gr}}

\end{center}

%\vfill

\vskip 1.0in \begin{abstract}

We  examine super symmetric representations of the $B$-type Hecke algebra. We exploit such representations to obtain new non-diagonal
solutions of the reflection equation associated to the super algebra $U_q(gl({\mathrm m}| {\mathrm n}))$. The boundary super algebra is
briefly discussed and it is shown to be central to the super symmetric realization of the $B$-type Hecke algebra \end{abstract}

\vfill \baselineskip=16pt

\end{titlepage}

\tableofcontents

\def\baselinestretch{1.2} \baselineskip 20 pt \no

\section{Introduction}

There exist various investigations regarding the solution of the reflection equation \cite{cherednik, sklyanin}, and although these
studies in the case of the usual Lie algebras and their $q$-deformed counterparts \cite{deve}--\cite{lima2} are rather exhaustive there
seem to exist gaps when one looks at the corresponding solutions associated to the super symmetric algebras. An exhaustive
classification of solutions of the reflection equation associated to a specific algebra is a fundamental physical and mathematical
issue, given that such investigations provide new boundary conditions that may for instance drastically alter the physical behavior as
well as the symmetry governing the corresponding system.

One finds numerous works presenting diagonal and non-diagonal solutions of the reflection equation associated to various super algebras.
More precisely, in \cite{annecy} diagonal and non-diagonal solutions are derived for the super Yangians $osp({\mathrm m}|{\mathrm n})$
and $gl({\mathrm m}|{\mathrm n})$, while in \cite{galleas1} non-diagonal solutions are obtained for the $sl(2|1)$ case (super symmetric
t-J model). In \cite{chin, ragoucy} purely diagonal solutions for the $U_q(gl({\mathrm m}|{\mathrm n}))$ are presented. Non-diagonal
solutions of the reflection equation associated to the deformed algebras $U_q(sl({\mathrm r}|2{\mathrm m})^{(2)})$ and $U_q(osp({\mathrm
r}|2{\mathrm m}))$ are presented in \cite{galleas}, whereas in \cite{zhang} a generic solution for the $U_q(sl(1|1))$ case is derived.
Here, we shall focus on the general $U_q(gl({\mathrm m}|{\mathrm n}))$ algebra and we shall find non-diagonal solutions of the
reflection equation. Note that this is the first time that non-diagonal solutions for the generic $U_q(gl({\mathrm m}|{\mathrm n}))$
case are derived, some relevant results are also presented in \cite{lima-santos}.

The main aim of the present investigation is the study of non-diagonal super symmetric representations of the B-type Hecke algebra
\cite{hecke, superhe, heckeb, heckeb1}. Using these representations we shall be able to identify novel generic non-diagonal solutions of the reflection
equation associated to $U_q(gl({\mathrm m}|{\mathrm n}))$. Moreover, based on these non-diagonal solutions we extract the associated boundary non-local
charges, which form the boundary super algebra and in the fundamental representation, are central to the super symmetric realization of
the $B$-type Hecke algebra.

The outline of this paper is as follows: in the subsequent section we introduce some basic notation on the super algebras and we also
recall the $U_q(\widehat{gl({\mathrm m}|{\mathrm n})})$ $R$-matrix and the corresponding super symmetric representations of the $A$-type
Hecke algebra. We then derive novel super symmetric representations of the $B$-type Hecke algebra and we identify the relevant
non-diagonal reflection matrices. In section 3 we discuss the boundary super algebras associated to the aforementioned reflection
matrix. We extract the boundary non-local charges, which form the boundary super algebra and are central to the $B$-type Hecke algebra.

\section{Deriving reflection matrices}

Our primary objective in this section is to derive new non-diagonal solutions of the reflection equation. This will be achieved by using
super symmetric representations of the $B$-type Hecke algebra, which will be defined subsequently. Both the representations as well as
the non-diagonal $K$-matrices are novel.

\subsection{Preliminaries}

Before we proceed presenting our main results it is necessary to introduce some useful notation associated to super algebras. Consider
the ${\mathrm m}+{\mathrm n}={\cal N}$ dimensional column vectors $\hat e_i$, with 1 at position $i$ and zero everywhere else, and the
${\cal N} \times {\cal N}$ $e_{ij}$ matrices defined as $(e_{ij})_{kl} = \delta_{ik} \delta_{jl}$. Then define the gradings: \be [\hat
e_i] = [i], ~~~~~[e_{ij}]= [i]+[j]. \ee The tensor product is graded as: \be (A_{ij}\otimes A_{kl}) (A_{mn}\otimes A_{pq}) =
(-1)^{([k]+[l])([m]+[n])} A_{ij} A_{mn} \otimes A_{kl} A_{pq}. \ee It is also convenient for what follows to introduce the distinguished
and symmetric grading, corresponding apparently to the distinguished and symmetric Dynkin diagrams. In the distinguished grading we
define: \begin{equation} [i]=\left\{ \begin{array}{ll}  0\,, & 1 \leq i \leq {\mathrm m}\,,\\ 			       1\,, & {\mathrm m}+1\leq
i \leq {\mathrm m}+{\mathrm n}\,.
  \end{array}\right.
\end{equation} In the $gl({\mathrm m}|2{\mathrm k})$ case we also define the symmetric grading as: \begin{equation} [i]=\left\{
\begin{array}{ll}  0\,, & 1 \leq i \leq {\mathrm k}\,, ~~~~~{\mathrm m}+{\mathrm k} +1 \leq i \leq {\mathrm m}+{2\mathrm k}\\ 			
1\,, & {\mathrm k}+1 \leq i \leq {\mathrm m}+{\mathrm k}\,.
  \end{array}\right.
\end{equation}

We shall focus here, as already mentioned, on the $U_q((\widehat{gl({\mathrm m}|{\mathrm n})})$ algebra. The $R$-matrix associated to
this algebra satisfies the graded Yang-Baxter equation \cite{baxter, kulishgr} \be R_{12}(\lambda_1 -\lambda_2)\ R_{13}(\lambda_1)\
R_{23}(\lambda_2) =R_{23}(\lambda_2)\ R_{13}(\lambda_1)\ R_{12}(\lambda_1 -\lambda_2), \ee and is given by the following expressions
\cite{perk, perk1}: \be R(\lambda) = \sum_{i=1}^{{\cal N}} a_i(\lambda)\ e_{ii} \otimes e_{ii}+ b(\lambda) \sum_{i\neq j=1}^{{\cal N}} e_{ii}
\otimes e_{jj} + \sum_{i\neq j=1}^{{\cal N}} c_{ij}(\lambda)\ e_{ij} \otimes e_{ji}, \label{rr1} \ee where we define \be a_j(\lambda) =
\sinh(\lambda +i\mu -2 i \mu [j]), ~~~~~ b(\lambda) = \sinh (\lambda), ~~~~c_{ij}(\lambda) =\sinh(i \mu) e^{sign(j-i)\lambda}
(-1)^{[j]}. \label{rr2} \ee It is clear that the $R$-matrix may be written in the following form \be R(\lambda) = e^{\lambda}R^+ -
e^{-\lambda} R^-. \ee Also the matrix $\check R = PR$ may be written in terms of super symmetric representations of the $A$-type Hecke
algebra; $P$ is the permutation operator and it has the following form \be P = \sum_{i,j } (-1)^{[j]} e_{ij} \otimes e_{ji}. \ee Indeed
we have (see also \cite{jimbo}): \be \check R_{i i+1}(\lambda) = \sinh (\lambda) U_i + \sinh(\lambda + i\mu), \ee where $U_i$ is a super
symmetric representation of the $A$-type Hecke algebra \cite{hecke, superhe}. We recall the $A$-type Hecke algebra $H_N(q)$ is defined
by generators ${\mathbb U}_i$, $\ i = 1, \ldots, N-1$, and exchange relations: \be && {\mathbb U}_i\ {\mathbb U}_{i+1}\ {\mathbb U}_i -
{\mathbb U}_i = {\mathbb U}_{i+1}\ {\mathbb U}_i\ {\mathbb U}_{i+1}-{\mathbb U}_{i+1} \label{a1}\\ &&        {\mathbb U}_i^2 = \delta\
{\mathbb U}_i   \label{a2} \\ &&   [{\mathbb U}_i,\ {\mathbb U}_j] =0, ~~~|i- j| >1 \label{a3} \ee where $\delta = -(q+q^{-1})$ and $q =
e^{i\mu}$.

Consider the following super symmetric representation \cite{superhe} associated to the $U_q(\widehat{gl({\mathrm m}|{\mathrm n})})$
$R$-matrix. Let \be && U = \sum_{a,b=1}^{{\cal N}} f_{ab}\ e_{ab}\otimes e_{ba} + \sum_{a, b=1}^{\cal N} t_{ab}\ e_{aa} \otimes e_{bb},
~~~~\mbox{where} \non\\ && f_{aa} = 0, ~~~~~f_{ab} =(-1)^{[b]} ~~~ a\neq b, \non\\ && t_{aa}= (-1)^{[a]}q^{1 -2 [a]} -q, ~~~~~ t_{ab} =
-q^{-sign(a-b)} ~~~a \neq b, \ee then we obtain the super symmetric representation $\pi:\ H_N(q) \hookrightarrow \mbox{End}(({\mathbb
C}^{{\cal N}})^{\otimes N})$ such that \be \pi({\mathbb U}_i)= U_i = {\mathbb I} \otimes \dots {\mathbb I} \otimes \underbrace{U}_{i,\
i+1} \otimes \ldots \otimes {\mathbb I}. \ee

\subsection{Reflection matrices from the $B$-type Hecke algebra}

Our aim now is to find super symmetric representations of the $B$-type Hecke algebra as candidate solutions of the super reflection
equation \cite{cherednik, sklyanin} \be R_{12}(\lambda_1 -\lambda_2) K_1(\lambda_1) R_{21}(\lambda_1 + \lambda_2) K_2(\lambda_2)=
K_2(\lambda_2)R_{12}(\lambda_1 + \lambda_2) K_1(\lambda_1) R_{21}(\lambda_1 -\lambda_2). \label{re} \ee where $R$ here is given by
(\ref{rr1}), (\ref{rr2}).

Recall the $B$-type Hecke algebra \cite{heckeb, heckeb1} $B_{N}(q, Q)$ is defined by generators ${\mathbb U}_i$ that satisfy
(\ref{a1})-(\ref{a3}) and ${\mathbb U}_0$ with: \be && {\mathbb U}_1\ {\mathbb U}_0\ {\mathbb U}_1 \ {\mathbb U}_0 - \kappa\  {\mathbb
U}_1 \ {\mathbb U}_0 = {\mathbb U}_0\ {\mathbb U}_1\ {\mathbb U}_0 \ {\mathbb U}_1 - \kappa\  {\mathbb U}_0 \ {\mathbb U}_1 \label{h2}\\
&& {\mathbb U}_0^2 = \delta_0\ {\mathbb U}_0 \label{h1}\\ && [{\mathbb U}_0,\ {\mathbb U}_i ] =0, ~~~~i>1 \label{h3} \ee where it is
convenient to parametrize as: $\delta_0= -(Q+Q^{-1})$ and $\kappa = qQ^{-1} + q^{-1} Q$.

Inspired essentially by the structure of the representations in the non super symmetric case \cite{kumu, myhecke, nichols} we consider
the following form for the non-diagonal super symmetric representation. Let \be {\mathrm e} = \sum_{a=1}^{{\cal N}} h_a\ e_{aa} +
\sum_{a=1}^{{\cal N}}  c_a\ e_{a \bar a} \label{ee} \ee where the parameters $h_a,\ c_a$ are a priori free. We define generally the
conjugate index $\bar a$ such that $[a] = [\bar a]$, and more specifically: \be && \bar a = 2{\mathrm k} +{\mathrm m} +1 -a;
~~~~\mbox{Symmetric diagram} \non\\ && \bar a = {\mathrm m} +1- a, ~~\mbox{$a$ bosonic}; ~~~\bar a = 2{\mathrm m}+{\mathrm n}+1-a,
~~\mbox{$a$ fermionic}; ~~~~ \mbox{Distinguished diagram.} \non\\ \ee Consider tensor type super symmetric representations of the
$B$-type Hecke algebra; $\pi: B_N(q, Q) \hookrightarrow \mbox{End}(({\mathbb C}^{\cal N})^{\otimes N})$ such that \be \pi({\mathbb U}_i)
&=& U_i={\mathbb I} \otimes {\mathbb I} \ldots \otimes \underbrace{U}_{\mbox{ $i,\ i+1$}} \otimes \ldots \otimes {\mathbb I} \non\\
\pi({\mathbb U}_0) &=& U_0= \underbrace{{\mathrm e}}_{\mbox{$1$}} \otimes\ {\mathbb I} \ldots \otimes {\mathbb I}. \label{rep} \ee

To prove that (\ref{ee}), (\ref{rep}) is a super symmetric representation although tedious is quite straightforward. We want  $U_1,\
U_0$ to satisfy the $B$-type Hecke algebraic relations. First from the constraint (\ref{h1}): \be U_0^2 &=& \sum_{a = 1}^{{\cal N}} h_a
(h_a + c_{a}c_{\bar a}h_a^{-1} )\ e_{aa} + \sum_{a=1}^{\cal N} c_a (h_a + h_{\bar a})\ e_{a \bar a}. \ee From the equation above and
(\ref{h1}) we conclude: \be && h_a = h_{\bar a}^{-1}, ~~~ h_a + h_{\bar a} = -(Q + Q^{-1}), ~~~c_a c_{\bar a} =1, \non\\ && \mbox{or}
~~~~h_a =h_{\bar a} =c_a =c_{\bar a} =0. \ee

Moreover, after some quite tedious algebra we conclude \be && U_0\  U_1\ U_0\ U_1 -\kappa\ U_0\ U_1 = \ldots \non\\ &&= \sum_{a, b}
(t_{ab} f_{ab} h_a^2 + t_{\bar a b}f_{ab} c_a c_{\bar a} + f_{ab}t_{ba}h_a h_b  - \kappa f_{ab} h_a)\ e_{ab} \otimes e_{ba} \non\\ && +
\sum_{a, b} (t_{ab} t_{ab}c_{\bar a} h_a  + f_{ab}f_{ba} c_{\bar a} h_b  (-1)^{[a] +[b]} + t_{\bar a b} t_{ab} h_{\bar a} c_{\bar a}
-\kappa c_{\bar a} t_{ab})\ e_{\bar a a} \otimes e_{bb} \non\\ && +\sum_{a,b} f_{ab} t_{\bar b a} h_a c_b\  e_{a \bar b} \otimes e_{ba}
\non\\ && +\sum_{ab} (f_{ab} t_{ba} c_{\bar a} h_b  + f_{ab} t_{\bar a b} c_{\bar a} h_{\bar a}  + f_{ab}t_{ab}c_{\bar a} h_a-\kappa
c_{\bar a} f_{ab})\ e_{\bar a b} \otimes e_{ba} \non\\ && +\sum_{a, b} f_{a b} t_{\bar b a} c_{\bar a} c_b\ e_{\bar a \bar b} \otimes
e_{ba} + \sum_{a, b}f_{ab} f_{\bar b a} c_{\bar a} c_b (-1)^{[a]+[b]}\ e_{\bar a a} \otimes e_{b \bar b} \label{1} \ee and \be && U_1\
U_0\ U_1\ U_0 -\kappa\ U_1\ U_0 = \ldots \non\\ && = \sum_{ab} (f_{ab} t_{\bar b a} c_b c_{\bar b} +t_{ab} f_{ab}  h_a h_b + f_{ab}
t_{ba} h_b^2 - \kappa f_{ab} h_b)\ e_{ab} \otimes e_{ba} \non\\ && + \sum_{a, b} (f_{\bar a b} f_{b \bar a} c_{\bar a} h_b(-1)^{[a]+[b]}
+ t_{\bar a b}t_{ab} c_{\bar a} h_a + t_{\bar a b}t_{\bar a b} h_{\bar a} c_{\bar a} -\kappa t_{\bar a b} c_{\bar a})\ e_{\bar a a}
\otimes e_{bb} \non\\ && +\sum_{a, b} (f_{ab}t_{ba}h_b c_b + f_{ab} t_{\bar b a} c_b h_{\bar b} + f_{ab} t_{ab} h_a c_b -\kappa f_{ab}
c_b )\ e_{a \bar b} \otimes e_{ba} \non\\ && +\sum_{a, b} f_{ab} t_{\bar a b} c_{\bar a} h_b\ e_{\bar a b} \otimes e_{ba} \non\\ && +
\sum_{a,b} f_{a b} t_{\bar a b} c_{\bar a} c_b\ e_{\bar a \bar b} \otimes e_{ba} + \sum_{a, b} f_{\bar a b} f_{\bar b \bar a} c_{\bar a}
c_b (-1)^{[a]+[b]}\ e_{\bar a a} \otimes e_{b \bar b}. \label{2} \ee Finally, after appropriately combining the terms of the equations
above we show that the quadratic constraint (\ref{h2}) is indeed satisfied provided that $\kappa = q Q^{-1} + q^{-1}Q$, and we can now
derive the explicit form of the non-diagonal solution.

More specifically, the generic solution we find is associated to the {\it symmetric Dynkin diagram} only,
due to technical specificities regarding the fermionic and bosonic indices and reads as follows:
\\
\\
{\bf Symmetric Dynkin diagram:}
\be
&& h_a = h_{\bar a}^{-1}=-Q^{-1}, ~~~~c_a c_{\bar a} =1,  ~~~~1 \leq a \leq L,
\non\\
&& h_a =  h_{\bar a} = c_a =c_{\bar a}=0, ~~~~ L < a \leq{{\mathrm m} +2{\mathrm k} \over 2}, ~~~~1 \leq L \leq
{{\mathrm m} +2{\mathrm k} \over 2}; \non\\
&& h_{{{\mathrm m} +2 {\mathrm k} +1 \over 2}}=0 ~~~~\mbox{if ${\mathrm m}$ odd}.
\ee
Notice that in this case the solution may have purely bosonic purely fermionic and mixed indices.

The super symmetric representation is then expressed as: \\ \\
{\bf Symmetric Dynkin diagram:} \be &&{\mathrm e}=
\sum_{a=1}^{L}(-Q^{-1} e_{aa}  -Q e_{\bar a \bar a}) + \sum_{a=1}^{L} (c_a e_{a \bar a} + c_{\bar a} e_{\bar a a}), ~~~~c_a c_{\bar a}
=1, \non\\ && 1 \leq L \leq {{\mathrm m} +2 {\mathrm k} \over 2}. \ee

Using the theorem proved in \cite{lema, lema1, doma, myhecke} we may express the $K$-matrix solution of the reflection equation as: \be
K(\lambda) &=& x(\lambda)\ {\mathbb I}  + y(\lambda)\ {\mathrm e}, \non\\ x(\lambda) &=& -{\delta_0 \over 2 i  \sinh (i \mu)}
\cosh(2\lambda + i\mu) - {\kappa \over 2 i \sinh (i \mu)} \cosh(2\lambda)- \cosh 2 i \mu \zeta, ~~~y(\lambda)=  i \sinh (2\lambda).
\non\\ \ee Let also $Q= i e^{i \mu m }$ then the non zero entries of the reflection matrix are written below:
\\ \\ {\bf Symmetric Dynkin diagram:} \be && K_{aa}(\lambda) = e^{2 \lambda} \cosh i\mu m - \cosh 2 i \mu \zeta, ~~~K_{\bar a \bar a}(\lambda)= e^{-2
\lambda} \cosh i\mu m - \cosh 2i \mu \zeta, \non\\ && K_{a \bar a}(\lambda)= i c_a \sinh 2 \lambda, ~~~ K_{\bar a a}(\lambda)= i c_{\bar
a} \sinh 2 \lambda, ~~~~1\leq a\leq L \non\\ && K_{aa}(\lambda) = K_{\bar a \bar a}(\lambda)= \cosh (2\lambda +i m \mu) - \cosh 2i \mu
\zeta, ~~~ K_{a\bar a}(\lambda)=  K_{\bar a a}(\lambda)=0, ~~~~L< a \leq {{\mathrm m} +2{\mathrm k} \over 2}; \non\\ && 1 \leq L \leq
{{\mathrm m} +2{\mathrm k} \over 2} \non\\ && K_{AA} = \cosh (2\lambda +i m \mu) - \cosh 2i \mu \zeta, ~~~A={{\mathrm m}+2{\mathrm k}
+1\over 2}~~~~ \mbox{if ${\mathrm m}$ odd}. \label{k} \ee
In the $q=1$ case we get non-diagonal solutions of the isotropic $gl({\mathrm
m}|{\mathrm n})$ case and the hyperbolic functions become rational. Also in the case that ${\mathrm n} =0$ we recover non super
symmetric solutions found in \cite{kumu, nichols}. Note also that very similar solutions (but with purely bosonic or fermionic indices)
are also recovered in \cite{karaiskos} for the {\it distinguished Dynkin diagrams}.

An interesting problem to pursue is the derivation of non-diagonal solutions associated to cyclotomic Hecke algebras, that is affine
Hecke type algebras with more generic constraints instead of the quadratic one (\ref{h1}) i.e. \be \sum_{k=1}^p \alpha_k {\mathbb
U}_0^{k} =0. \ee The latter is a significant question even in the context of $U_q(gl (n))$ algebras (see e.g. \cite{kumu, nichols}),
especially when one analyzes the boundary behavior of $A_{n-1}^{(1)}$ affine Toda field theories \cite{doikouaf, doikouaf1}. In this case in
particular as discussed in \cite{doikouaf, doikouaf1} the most general solution possible is needed in order to eliminate certain discrepancies
arising at the classical Hamiltonian formalism. An exhaustive classification of the solutions of the (super) reflection equation based
on the cyclotomic Hecke algebra will be pursued elsewhere.

\section{The boundary super algebra; central elements}

We shall now briefly discuss the so called boundary super algebra, which turns out to be central to the super symmetric realization of
the $B$-type Hecke algebra. Let us first recall the associated double row transfer matrix \cite{sklyanin} \be t(\lambda) =str_0 \Big
\{M_0\ K_0^{(L)}(\lambda)\ {\mathbb T}_0(\lambda) \Big \}. \label{transfer} \ee ${\mathbb T}$ is a tensor representation of the
reflection algebra \be {\mathbb T}_0(\lambda) = T_0(\lambda)\ K_0(\lambda)\ \hat T_0(\lambda), ~~~~~\hat T(\lambda) = T^{-1}(-\lambda),
\ee where $T$ is the monodromy matrix defined as \cite{tak, fad, kusk}: \be T_0(\lambda)= R_{0N}(\lambda)\ R_{0 N-1}(\lambda) \ldots
R_{02}(\lambda)\ R_{01}(\lambda). \ee Also the matrix $M$ is defined in the $U_q(gl({\mathrm m}|{\mathrm n}))$ case as: \be M =
\sum_{k=1}^{{\cal N}} q^{{\cal N} -2 k +1} q^{-2[k] + 4 \sum_{i=1}^k [i]} e_{kk}. \ee It is also shown, using the fact that ${\mathbb
T}$ and $K^{(L)}$ satisfy the reflection equation, that \cite{sklyanin} \be [t(\lambda),\ t(\mu)] =0, \ee ensuring the integrability of
the associated system. We consider here for simplicity the left boundary $K^{(L)} \propto {\mathbb I}$.

Due to the fact that the $R$-matrix reduces to the permutation operator for $\lambda =0$, a local Hamiltonian may be deduced from
(${\cal H} \propto {d t(\lambda) \over d \lambda}|_{\lambda =0}$), which can be expressed in terms of super symmetric representations of
the $B$-type Hecke algebra (for more details see e.g.~\cite{doma, myhecke}). Also, from the asymptotic behavior of ${\mathbb T}$, by
keeping the leading contribu tion i.e. ${\mathbb T}(\lambda \to \infty) \sim {\mathbb T}^{\pm} +{\cal O}(e^{{\mp} 2\lambda})$, we obtain
the so called boundary super algebra. More precisely, the elements ${\mathbb T}_{ab}^{\pm}$ are the boundary non-local charges, which
form the boundary super algebra -- a non-abelian algebra in general-- with exchange relations dictated by: \be R_{12}^{\pm}\ {\mathbb
T}^{\pm}_1\ R_{21}^{\pm}\ {\mathbb T}_2^{\pm}= {\mathbb T}_2^{\pm}\ R_{21}^{\pm}\ {\mathbb T}^{\pm}_1\ R_{12}^{\pm}. \ee In this
context, it may be shown that the boundary super algebra is central to the super symmetric realization of the $B$-type Hecke algebra
i.e. (the proof goes along the same lines as in \cite{doikoutwin}): \be \big [U_{i},\ {\mathbb T}^{\pm}_{ab} \big ]=0, \qquad \big
[U_0,\ {\mathbb T}^{\pm}_{ab} \big ]=0,  \qquad i\in \{1, \dots,  N-1 \}, ~~~~a,\ b \in \{1, \ldots, {\cal N}\}, \label{a} \ee where
apparently ${\mathbb T}_{ab}^{\pm} \in \mbox{End}(({\mathbb C}^{\cal N})^{\otimes N})$. The boundary super algebra also turns out to be
an exact symmetry of the double row transfer matrix, however the explicit proof as well as a detailed discussion will be presented
elsewhere (see \cite{mynew}).

It is worth mentioning that as in the case of the non super symmetric quantum algebras appropriate choice of boundary conditions leads
to known quantum algebras as boundary symmetries. For instance, if  we choose $K \propto {\mathbb I}$ then one may show that the
boundary algebra is the $U_q(gl({\mathrm m}| {\mathrm n}))$ (see \cite{mynew} for details on the proof), which apparently is central to
the super symmetric realization of the $A$-type Hecke algebra  ($U_0$ in this case is trivial).

\end{document}